\documentclass[preprint,12pt]{elsarticle}

\usepackage[svgnames]{xcolor}
\usepackage{crayola}
\usepackage{verbatim}
\usepackage{amsmath}
\usepackage{amsthm}
\usepackage{mycal}
\usepackage{mybold}
\usepackage{mymathbb}
\usepackage{changepage}
\usepackage{graphicx}            

\usepackage{gastex}
\usepackage{float}
\usepackage{tikz}
\usepackage{pgfplots}
\usepackage{tikz-inet} 
\usetikzlibrary{shapes,decorations,arrows,automata,plotmarks,matrix,decorations.pathmorphing}

\usepackage{graphicx}		
\usepackage{hyperref}			
\usepackage{array}   
\usepackage{pict2e}  
\usepackage[english]{babel}	
\usepackage{listings}

\usepackage{stmaryrd}
\usepackage{amsfonts}
\usepackage{booktabs}
\usepackage{amssymb}
\usepackage{algorithmicx}
\usepackage{algorithm}
\usepackage[noend]{algpseudocode}
\usepackage{bm}
\usepackage{subcaption}
\usepackage{diagbox}

\usepackage[T1]{fontenc}     
\usepackage[utf8]{inputenc}  

\algnewcommand\algorithmicinput{\textbf{Input:}}
\algnewcommand\INPUT{\item[\algorithmicinput]}

\newtheorem{definition}{Definition}
\newtheorem{theorem}{Theorem}
\newtheorem{proposition}[theorem]{Proposition}

\newtheorem{corollary}[theorem]{Corollary}

\definecolor{yel}{cmyk}{0.1,0.95,1.0,0.0}
\definecolor{head}{cmyk}{0.1,1.0,0.8,0.15}
\definecolor{vector}{cmyk}{0.0,0.8,1.0,1.0} 
\definecolor{tan}{cmyk}{0.30,0.50,0.60,0}
\definecolor{orange}{cmyk}{0.0,0.6,1.0,0.1}
\definecolor{emp}{cmyk}{1.0,0.5,0.0,0}       
\definecolor{names}{cmyk}{1.0,0.0,1.0,0.14}
\definecolor{pink}{cmyk}{0.0,0.8,0,0}
\definecolor{paleyellow}{cmyk}{0,0,0.6,0.0}
\definecolor{darkyellow}{cmyk}{0,0.2,1.0,0.2}

\def\normq{{\sf q}}
\def\eqdef {:=}    

\def\q {$\kern15pt$}    
\def\?{\discretionary{}{}{}}  

\def\vol{{\rm vol}}

\def\fraku{\mathfrak{u}}

\newtheorem{rmk}[theorem]{Remark}

\definecolor{darkspringgreen}{rgb}{0.09, 0.45, 0.27}

\newif\ifnotes\notestrue

\newcommand{\modified}[1]{{#1}}

\newcommand{\hpl}[1]{{}}

\def\hmarion#1{}

\def\hmaxime#1{}

\begin{document}

\begin{frontmatter}

\title{An algorithm to compute the $t$-value of a digital net and 
  of its projections}

\author[POLY]{Pierre Marion}
\ead{pierre.marion@polytechnique.edu}
\address[POLY]{CMAP, Ecole Polytechnique, 1 route de Saclay, 91128 Palaiseau, France}

\author[POLY]{Maxime Godin}
\ead{maxime.godin@polytechnique.edu}

\author[UM]{Pierre L'Ecuyer\corref{cor}}
\cortext[cor]{Corresponding author. \texttt{\tt lecuyer@iro.umontreal.ca}}
\ead{lecuyer@iro.umontreal.ca}
\address[UM]{DIRO, Universit\'e de Montr\'eal, Pavillon Aisenstadt, 
    C.P. 6128, Succ. Centre-Ville, Montr\'eal, H3C 3J7, Canada}

\begin{abstract} 
Digital nets are among the most successful methods to construct low-discre\-pancy point sets for quasi-Monte Carlo integration.  
Their quality is traditionally assessed by a measure called the $t$-value. 
A refinement computes the $t$-value of the projections over subsets of coordinates and takes a weighted average (or some other function) of these values. 
It is also of interest to compute the $t$-values of embedded nets obtained by taking subsets of the points. 
In this paper, we propose an efficient algorithm to compute such measures and we compare our approach with previously proposed methods both empirically and in terms of computational complexity.  
\end{abstract}


\end{frontmatter}

\section{Introduction}
\label{sec:introduction}

The \emph{Monte Carlo method} (MC) simulates random variables from arbitrary distributions by 
uniform sampling of independent random points from the $s$-dimensional unit hypercube $[0,1)^s$ for some integer $s$, 
followed by an appropriate transformation to achieve the target distribution \cite{sASM07a}.
\emph{Quasi-Monte Carlo} (QMC) replaces these independent points by deterministic point sets that are carefully constructed to cover the hypercube much more evenly than i.i.d. points 
\cite{rDIC10a,vLEC09f,vLEC18a,rNIE92b,rNIE05a}.
The most popular construction methods are lattice rules and digital nets.
This paper concerns the latter.

An $s$-dimensional \emph{digital net in base $b$} is a construction that defines $n = b^k$ points 
in $[0,1)^s$ for some integer $k > 0$, via a linear mapping, as detailed in Section~\ref{sec:digital-net}. 
A traditional measure of quality for these constructions is the so-called $t$-value,
which measures in some sense the lack of uniformity of the points.
Bounds on the integration error in terms of the $t$-value of the net have been derived,
as well as bounds on the variance when the digital net is randomized in specific ways \cite{rDIC10a,rNIE92b,vOWE97b}.
More refined error or variance bounds can be obtained if we rewrite the integrand as a sum of lower-dimensional
components, e.g., via an ANOVA decomposition, and bound the error or variance of each term.
Doing this using $t$-values involves bounding a weighted sum of the $t$-values of the $2^s-1$ projections 
of the point set over subsets of coordinates.  
This motivates \textit{figures of merit} defined by such sums, to evaluate the quality of 
digital nets and select good ones for QMC integration.
To compute these figures, one must compute the $t$-value for each projection, 
which may take excessive computing time, given that calculating a single $t$-value already requires
testing the rank of a combinatorial number of matrices.

It is also useful to construct and test sequences of \textit{embedded digital nets} for which each net
is a subset of the next one and such that each net in the sequence has good quality. 
Generally, if the cardinality of a net is $b^k$, the cardinality of the next one will be $b^{k+1}$.
One then wants to compute the figure of merit for each net in the sequence, and perhaps take
the worst-case as a global measure of quality of the sequence.  
This is useful when the required number of points is unknown in advance and one wishes to increase 
$n$ until reaching the desired accuracy \cite{rDIC10a,vHIC01a}.

The aim of this paper is to propose a new algorithm to compute the $t$-value
of a given digital net, or the $t$-values of an arbitrary choice of its projections on subsets of coordinates,
and do this for sequences of embedded nets as well.
Our algorithm outperforms previously known methods when $n$ is large 
or when we want the $t$-values of many projections.
We provide an open-source software that implements our algorithm as well as three 
previously-proposed ones from \cite{rDIC13a,rPIR01a,rSCH99a}, and we provide a detailed 
comparison in terms of both empirical performance and computational complexity.
For now, the software is implemented only for basis $b=2$, which is by far the most widely used
basis in practice, for efficiency reasons.

In Section~\ref{sec:digital-net}, we recall the notions of digital net, $t$-value, 
as well as error bounds and figures of merit based on $t$-values.
The new algorithm is introduced in Section~\ref{sec:algo}.
A time complexity comparison is presented in Section~\ref{sec:complexity}.
In Section~\ref{sec:numerical}, we present empirical performance comparisons.

\section{Digital nets and $t$-values}
\label{sec:digital-net}

\subsection{Definitions}

Let $b \geq 2$ be a prime number and let $\FF_b$ denote the finite field of cardinality $b$. 
We identify the elements of $\FF_b$ with $\{0,1,\dots,b-1\}$ 
where 1 is the unit element and the operations correspond to the arithmetic modulo $b$.
Let $s \geq 1$, $k \geq 1$ and $r \geq k$ be integers. Let $C_1,\dots,C_{s}$ be $s$ matrices of size $r\times k$ over $\FF_b$. For $i=0,\dots,b^k-1$, let $\bm{i} = (a_0, \dots, a_{k-1})^T \in\FF_b^k$ be the column vector that contains
the $k$ digits of the base-$b$ expansion of $i = \sum_{\ell=0}^{k-1} a_\ell b^\ell$.  
For each coordinate $j=1,\dots s$, let
\[
  (y_{i, j, 1}, \dots, y_{i, j, r})^T = C_j \cdot \bm{i} 
\quad\mbox{and}\quad
  x_{i, j} = \sum_{\ell=1}^{r} y_{i, j, \ell}b^{-\ell}.
\] 
and let  $\bm{x}_i = (x_{i, 1}, \dots, x_{i, s})^T$. 
The point set $\cP = \{\bm{x}_0, \dots, \bm{x}_{n-1}\}$ is called a \emph{digital net in base $b$}. It contains $n = b^k$ points laying in the half-open $s$-dimensional unit hypercube $[0,1)^s$. The matrices $C_1,\dots,C_{s}$ are called the \emph{generator matrices} of $\cP$.
Niederreiter \cite{rNIE87b,rNIE92b} introduced the notion of digital net with a
more general definition, in which the operations are in a general finite \modified{commutative ring} and an 
arbitrary bijection can be used between the elements of \modified{this ring} and the ring $\ZZ_b$. 
Popular constructions that are special cases of the above definition were introduced 
earlier by Sobol' \cite{rSOB67a} (in base 2) and Faure \cite{rFAU82a} 
(in prime base $b\ge 2$).

In our definition, each coordinate has $r$ digits, 
but only the first $k$ digits will matter for the properties 
examined in this paper, so we can assume 
that $r=k$ and ignore the extra rows of the generator matrices.
We also always assume that the first $k$ rows of $C_j$ are linearly independent.  
Then, the digital net is \emph{fully projection-regular}, which means that for each $j$
the one-dimensional projection of $\cP$ over the $j$th coordinate
is the set $\{0, 1/n, \dots, (n-1)/n\}$.
These numbers are enumerated in a different order for the different coordinates,
i.e., the matrices $C_j$ implement permutations of this set,
and the choice of these permutations is crucial for the uniformity of the projections of 
$\cP$ over subspaces in two or more dimensions.

This uniformity of $\cP$ can be assessed as follows \cite{rDIC10a,vLEC02a,rNIE92b}.
We can select $s$ non-negative integers $q_1,\dots,q_s$ such that $q \eqdef q_1 + \cdots + q_s \le k$,
and divide axis $j$ in $b^{q_j}$ equal parts, for $j=1,\dots,s$.
This defines a partition of $[0,1)^s$ into $b^{q}$ identical rectangular boxes
which are shifted copies of $[0,b^{-q_1}) \times \cdots\times [0,b^{-q_s})$.
If each of those boxes contains exactly the same number of points from $\cP$, 
which has to be $b^{k-q}$ points in each box, we say that 
$\cP$ is \emph{$(q_1,\dots,q_s)$-equidistributed in base $b$}.

The digital net $\cP$ is said to be a \emph{$(t,k,s)$-net in base $b$}
if it is $(q_1,\dots,q_s)$-equidistributed for any $(q_1, ..., q_s)$ such that $q = q_1 + \cdots + q_s \le k-t$.
The \emph{$t$-value} of a digital net 
is the smallest $t$ for which it is a $(t,k,s)$-net.
We denote it by $t(\cS)$ where $\cS = (C_{1},\dots,C_{s})$ to emphasize that 
it is a function of the vector of generator matrices.
We want the $t$-value to be as small as possible, ideally 0, although it is known that 
$t=0$ is impossible unless ${b \ge s-1}$ \cite[Corollary 4.21]{rNIE92b},
and much tighter lower bounds on the best possible $t$ for any given values of
$b$, $k$, and $s$ can also be computed via linear programming \cite{mMAR00a}.

If $\fraku = \{j_1,\dots,j_d\} \subseteq \{1,\dots,s\}$ is a nonempty subset of coordinates, 
the \emph{projection} $\cP_{\fraku}$ of $\cP$ over the coordinates in $\fraku$ is also a
digital net, in $d=|\fraku|$ dimensions, with vector of generator matrices 
$\cS_{\fraku} = (C_{j_1},\dots,C_{j_d})$.
We are also interested in computing the $t$-values $t(\cS_{\fraku})$
for these projected point sets $\cP_{\fraku}$.
The assumption of full projection-regularity implies that $t(\cS_{\fraku})=0$
whenever $\fraku = \{j\}$, i.e., when the projection is over a single coordinate.

A sequence of embedded digital nets of cardinalities $b^{m_0}, b^{m_0+1},\dots, b^{k}$,
for two arbitrary integers $1 \le m_0 \le k$, can be obtained as follows. 
Let $\cS = (C_{1},\dots,C_{s})$ denote the vector of generator matrices of the largest net, with $m= k$.
For $j=1,\dots,s$ and $m = m_0,\dots,k$, let $C_{j,m}$ 
be the $k\times m$ matrix formed by the first $k$ rows and $m$ columns of $C_j$.
Then $\cS_m = (C_{1,m},\dots,C_{s,m})$ 
is the vector of generator matrices of a digital net which
is the restriction of the largest net to its first $b^m$ points. 
Each $C_{j,m}$ is assumed to be non-singular 
so that this digital net is also fully projection-regular.
We denote by $\cS_{\frak u, m}$ the vector of generator matrices 
for the projection $\frak u$, restricted to their first $k$ rows and $m$ columns.
In this context, we are interested in computing the $t$-values $t(\cS_{\frak u, m})$ 
for all $m \in \{m_0, \dots, k\}$ and a large set of projections $\frak u$.

\subsection{Figures of merit based on $t$-values of projections}

For a given vector of generator matrices $\cS$, we are interested in computing figures of merit of the following general form:
\begin{align}
  D_{\tilde{t}, \gamma}^{(\normq)}(\cS) 
	 &= \left[ \sum_{\emptyset \neq \mathfrak{u} \subseteq \{1, \dots, s\}}
   \left( \gamma_\mathfrak{u} \,\tilde{t}(\cS_{\mathfrak u})\right)^\normq \right]^{1/\normq} 
	 &\text{ if } \normq < \infty,
	\label{eq:fig-merit} \\
  D_{\tilde{t}, \gamma}^{(\infty)}(\cS) 
	 &= \max_{\emptyset \neq \mathfrak{u} \subseteq \{1, \dots, s\}}
   \gamma_\mathfrak{u} \,\tilde{t}(\cS_{\mathfrak u})
  &\text{ if } \normq = \infty,
	\label{eq:fig-merit-infinity}
\end{align} 
 where 
$\normq \in [1,\infty]$, 
$\gamma = \{\gamma_\mathfrak{u},\; \emptyset \neq \mathfrak{u} \subseteq \{1, \dots, s\}\}$ 
is a set of non-negative weights given to the projections \cite{vDIC04a} and $\tilde{t}$ is 
a real-valued function of the $t$-value and other parameters of the net (some examples are given below).

For two integers $1 \leq m_0 \leq k$ and a vector $\cS$ of generator matrices of size $k \times k$, 
we also want to compute the figure of merit for embedded nets:
\begin{align}	
  D_{\tilde{t}, \gamma, m_0:k}^{(\normq)}(\cS) 
	 &= \max_{m_0 \leq m \leq k} D_{\tilde{t}, \gamma}^{(\normq)}(\cS_m).
	\label{eq:fig-merit-embedded}
\end{align}
These general forms leave a lot of freedom for the choice of the function $\tilde{t}$ and the weights $\gamma$.
Certain choices of $\tilde{t}$, $\gamma$ and $\normq$ in (\ref{eq:fig-merit}) or (\ref{eq:fig-merit-infinity}) yield bounds on the integration error for a large class of integrands;
see Subsection~\ref{sec:error-bounds}.
The weights $\gamma$ should reflect the importance of having good uniformity in the corresponding projections:
the more relevant the projections, the greater the weights.  
This is discussed in \cite{rDIC10a}
for digital \modified{nets} and studied in \cite{vLEC12b,vLEC12a,vLEC16a} for lattice rules. 
The choice of $\normq$ determines the norm used to combine the values for the different projections. 
This choice does not affect our algorithm. 
The most common values are $\normq=2$ and $\normq=\infty$.  
The max in (\ref{eq:fig-merit-embedded}) can also be replaced by a weighted sum or a more general norm.

\subsection{Discrepancy-based error bounds}
\label{sec:error-bounds}

An important special case of the figure of merit (\ref{eq:fig-merit-infinity}) gives a bound on the 
\emph{weighted star discrepancy} of $\cP$, which in turns provides a bound on the integration error
for a large class of integrands, via the Koksma-Hlawka inequality \cite{rDIC10a,rLAR03a}.
For any $\bz \in[0,1]^s$, let $\bz_{\fraku}$ be the projection of $\bz$ on
the coordinates in $\fraku$ and $(\bz_\fraku,\bone)$ be the point $\bz$ in which the coordinates whose
indices are not in $\fraku$ have been replaced by 1.
For positive weights $\gamma_{\fraku}$, we can define the \emph{weighted Hardy-Krause variation} 
of a function $f : [0,1)^s\to \RR$ by 
\begin{equation}
\label{eq:HK}
  V_{{\rm HK},\gamma}(f) 
	  = \sum_{\emptyset\not=\fraku\subseteq\{1,\dots,s\}} \gamma_{\fraku}^{-1} \int_{[0,1]^{|\fraku|}} 
	   \left|\frac{\partial^{|\fraku|}}{\partial\bz_{\fraku}} f(\bz_{\fraku},\bone)\right| d\bz_{\fraku},
\end{equation}
assuming that the partial derivatives exist and are integrable. 
Moreover, when the integral is 0 for some $\fraku$, one can take $\gamma_{\fraku} = 0$.
We also define the \emph{weighted star-discrepancy} of $\cP$ as 
\[
  D^*_{n,\gamma}(\cP) = \sup_{\bz\in[0,1]^s} \max_{\emptyset\not=\fraku\subseteq\{1,\dots,s\}} \gamma_{\fraku}
    \left|\vol[\bzero_{\fraku},\bz_{\fraku}) - \frac{|\cP_{\fraku}\cap [\bzero_{\fraku},\bz_{\fraku})|}{n}\right|,
\]
where $[\bzero_{\fraku},\bz_{\fraku})$ is the box $[\bzero,\bz)$ projected to the coordinates in $\fraku$,
$\vol[\bzero_{\fraku},\bz_{\fraku})$ is the volume of that box,
and $|\cP_{\fraku}\cap [\bzero_{\fraku},\bz_{\fraku})|/n$ is the proportion of the points that are in that box.

The generalized \emph{Koksma-Hlawka inequality} \cite{mSLO98a} states that
\begin{equation}
  \left| \int_{[0, 1]^s} f(\bz) d \bz - \frac{1}{n} \sum_{i=0}^{n-1} f(\bx_i) \right| 
	\leq  V_{{\rm HK},\gamma}(f) \cdot D^*_{n,\gamma}(\cP).              
\label{eq:kokla}
\end{equation}
The weighted star discrepancy can in turn be bounded as follows 
\cite[\modified{Theorems 5.12 and 5.26}]{rDIC10a}:
\begin{align}
  D^*_{n,\gamma}(\cP)	& \leq D_{\tilde{t}, \gamma, k}^{(\infty)}(\cS)
\end{align}  
with
\begin{align}
  \tilde{t}(\cS_\fraku) 
	 &=  b^{t(\cS_\fraku)-k} \sum_{\ell=0}^{|\fraku|-1} {k-t(\cS_\fraku) \choose \ell} \modified{(b-1)^\ell}
\end{align}  
where $b$ is the base of the digital net and $k$ the size of the generator matrices in $\cS_\fraku$.
As another special case, \cite{rJOE08a} made searches for good direction numbers for Sobol' \modified{sequences}
using a heuristic figure of merit that turns out to be a special case of (\ref{eq:fig-merit-embedded}) with 
\[
  \tilde{t}(\cS_\fraku) =  \frac{t(\cS_\fraku)^p}{k-t(\cS_\fraku)+1}
  \text{ and }  
 \gamma_\mathfrak{u} = \left\{
    \begin{array}{ll}
        0.9999^{\min(j_1,j_2)-1} & \text{ if } \mathfrak{u} = \{j_1,j_2\},\\
        0 & \text{ otherwise,}
    \end{array}
  \right.
\]
where $k$ denotes the size of the generator matrices in $\cS_\fraku$  and $p > 0$ is a real-valued parameter.
Only the two-dimensional projections matter for this measure, and those whose coordinates
are both large matter less.

\hpl{
\begin{rmk}	\label{rmk:zero-weights}
In the next part, to speed up the computations, we will want to take some weights equal to zero. In this case, the previous theorem is still valid at the expense of restricting the integrand to a smaller space of functions: in (\ref{eq:sobolev}), we see that for each projection $\mathfrak{u}$, $\gamma_\mathfrak{u}$ can be taken zero if and only if $\int_{[0, 1]^{|\mathfrak{u}|}} \left| \frac{\partial^{|\mathfrak{u}|}}{\partial z_\mathfrak{u}} f(z_\mathfrak{u}, 1) \right| dz_\mathfrak{u} = 0$.
\end{rmk}
}

\section{Computing \modified{$t$}-values for digital nets and their projections}
\label{sec:algo}

Computing the $t$-value is a combinatorial problem: one must verify the equidistribution for 
a large number of possibilities of $(q_1,\dots,q_s)$, especially if $k$ and/or $s$ are large.
Computing the $t$-value for all or many projections, and for several embedded subnets, as required in (\ref{eq:fig-merit}), (\ref{eq:fig-merit-infinity}), and (\ref{eq:fig-merit-embedded}), is even more challenging. 

When $s$ is large, it can be impractical to use positive weights for all the $2^s-1$ non-empty
subsets $\fraku$ of coordinates.  In the rest of the paper, we assume \emph{finite-order weights} 
of maximal order $d_{\max} \le s$, 
which means that $\gamma_{\fraku} = 0$ whenever $|\fraku| > d_{\max}$ \cite[page 95]{rDIC10a}. Note that when $d_{\max} = s$, there is actually no
restriction at all on the weights.
The choice of $d_{\max}$ is related to the computation budget: the higher the budget, 
the larger $d_{\max}$ can be.

In this section, we start by explaining the reduction of the computation of the $t$-value to a simpler combinatorial linear algebra problem, linked to the determination of the ranks of a large number of matrices. After reviewing how this problem has already been tackled in the literature, we describe our proposed method. First we focus on the non-embedded setting, starting from the computation of the $t$-value of the net and then generalizing to the computation of the $t$-value over many projections. Finally, we extend our method to an embedded setting where we aim at computing the $t$-values of many projections for embedded nets.


\subsection{Computing the linear independence parameter and the t-value of the net}
\label{subsec:t-value}

Given a digital net $\cP$ of $b^k$ points with vector of generator matrices $\cS = (C_1,  \dots\modified{,} C_s)$, we aim at computing $t(\cS)$. In other words,  we search for the largest $q$ such that $\cP$ is $(q_{1},\dots,q_{s})$-equidistributed for all tuples $(q_{1},\dots,q_{s})$ of non-negative integers that sum to $q$. 
This $q$ is denoted $\rho(\cS)$, and is called the 
\textit{linear independence parameter} of the vector of generator matrices $\cS$.
We have the relation $t(\cS) = k - \rho(\cS)$.

As detailed in \cite{rDIC10a}, $\cP$ is $(q_{1},\dots,q_{s})$-equidistributed if and only if the $q\times k$ \textit{composition matrix} 
\[
 C =
\left[
\begin{array}{c}
  C_{1}^{(q_1)} \\
  \vdots \\
  C_{s}^{(q_s)}
\end{array}
\right]
\]
formed by 
the first $q_{1}$ rows of $C_{1}$, 
the first $q_{2}$ rows of $C_{2}$, \dots, 
the first $q_{s}$ rows of $C_{s}$, has full rank $q$.
To find $\rho(\cS)$, we need to find the largest value of $q$ for which all
the composition matrices of size $q \times k$ have full rank $q$. 
For any given $q$, there are $\binom{q+s-1}{s-1}$ such composition matrices \cite{iKLI82a}. 
Thus, in the worst case, we have to compute the rank of 
$\sum_{q = 1}^k \binom{q+s-1}{s-1}$ matrices. 
We can do this either by increasing or decreasing values of $q \in \{1, \dots, k\}$,
and stop as soon as we find $\rho(\cS)$.
The best choice of direction depends on the actual value of $t$.
Proposition \ref{prop:comp-t-value} gives exact expressions for the number of vector additions
for each direction.

The algorithms reported in the literature compute the ranks of the
composition matrices independently 
(without reusing the computations from one composition matrix to another),
until the $t$-value can be determined. 
For the special case of base $b = 2$, Schmid \cite{rSCH99a} visits the composition
matrices by increasing value of $q$.  For each composition matrix, he visits
all linear combinations of the rows to check \modified{if} they are nonzero, 
by using a Gray code so only one row is changed at each step.
Pirsic and Schmid \cite{rPIR01a} 
proposed a different method that applies Gaussian elimination \cite{mAGA17a} 
independently to each composition matrix to compute its rank. 
They found that this second method is typically faster than the first for large $b$
and when $k-t$ is not too large, and the first method was usually faster otherwise.
We will denote these two methods by S and PS, respectively. 

These two approaches are fine to compute a single $t$-value, but they do not scale well when dealing with the figures of merit 
in (\ref{eq:fig-merit}) to (\ref{eq:fig-merit-embedded}).  The latter involve
several \modified{$t$}-values, one for each of $\sum_{d = 1}^{d_{\max}} \binom{s}{d}$ projections $\mathfrak u$ such that $|\mathfrak u| \leq d_{\max}$. 
All in all, we have to compute the rank of up to $\sum_{d = 1}^{d_{\max}} \sum_{q = 1}^k \binom{s}{d}  \binom{q+d-1}{d-1}$ matrices. 
For $k = 10$ and $s = d_{\max}  = 5$, this represents over $1.2 \times 10^{4}$ matrices, cumulating $9.1 \times 10^{4}$ rows. 
In use cases, $k$ and $s$ easily go up to 20. 
In this case, the number of matrices equals $2.6 \times 10^{14}$, cumulating $4.9 \times 10^{15}$ rows. 
In the following, we explain how to optimize such a daunting computation, first on a single projection (optimizing the sum over $q$), 
then on many projections (optimizing the sum over $d$).

Our strategy takes advantage from the fact that one can enumerate the set 
of composition matrices in a way that 
any two consecutive composition matrices differ by exactly one row. 
This relies on the enumeration of weak compositions as proposed in \cite{iKLI82a}. 
We build upon the algorithm of Pirsic and Schmid \cite{rPIR01a}, but instead of computing from scratch the Gaussian elimination for each composition matrix, we reuse computations. 
We detail this algorithm in the following subsection.

\subsection{Efficient computation of the rank of all composition matrices}
\label{subsection:gauss}

We now explain the reduction used to compute the rank of the composition matrices, which is weaker than the reduction in row echelon form \cite{mAGA17a}.
We call it a \emph{reduction in almost row echelon form} (RAREF).
Then, we describe how a RAREF can be efficiently computed for a matrix $C'$ which differs by exactly one row from a matrix $C$ for which a RAREF is already known.

\begin{definition}
	\label{def:almost-reduced-row-echelon-form}
	In $\FF_b$, given a matrix C of size $q \times k$ \modified{such that $q \leq k$, a triplet $(L, T, \bm{p})$ where $L$ is a $q \times q$ matrix, $T$ is a $q \times k$ matrix and $\bm{p}$ is a subset of $\{1, \dots, k\}$, is said to be a RAREF of $C$ if the following properties hold:}
	\begin{itemize}
		\item[(a)] \modified{$L$ is a non-singular matrix such that $L \cdot C = T$;}
		\item[(b)] \modified{The columns of $T$ whose indexes are in $\bm{p}$ are distinct vectors of the canonical base of $\FF_b^q$. The nonzero coefficients of these columns are called \emph{pivots};}
		\item[(c)]  \modified{The cardinality of $\bm{p}$ equals the rank of $T$.}
	\end{itemize}
\end{definition}

\begin{algorithm}
\vskip 5pt
Let $L$ be the identity matrix of size $q \times q$, $T = C$ \modified{and $\bm{p} = \emptyset$.}

For each row $i$ of $T$, from the first to the last:
\begin{enumerate}
\item Use transvections to zero all the elements of the $i$th row of $T$ which \modified{belong to columns in $\bm{p}$}. Apply the same transvections to $L$.
\item Locate the first nonzero coefficient of the $i$th row of $T$, say in the $j$th column of $T$, and multiply this row so that this leading nonzero coefficient equals 1. This coefficient is a new pivot. \modified{Let $\bm{p} = \bm{p} \cup \{j\}$.} Apply the same multiplication to $L$. If such a coefficient does not exist, skip the row.
\item Otherwise, use transvections to zero all the elements of the $j$th column of $T$ other than the pivot. Apply the same transvections to $L$.
\end{enumerate}

Return \modified{$(L, T, \bm{p})$}.
\caption{Computing a RAREF}
\label{alg:pivoting}
\end{algorithm}

Given a matrix $C$, one finds \modified{a RAREF} by applying row operations on each row $i$ of $C$ from the first to the last, as described in Algorithm~\ref{alg:pivoting},
which is a variation of the Gauss-Jordan elimination \modified{\cite{mAGA17a}}. 
We refer to this as \emph{pivoting}.
%
Since $L$ is non-singular, the rank is invariant \modified{to} a left-multiplication by $L$,
so the rank of $C$ must equal the number of pivots in $T$\modified{, i.e., the cardinality of $\bm{p}$.}
\modified{The following proposition proves the correctness of Algorithm~\ref{alg:pivoting}.

\begin{proposition}
\label{prop:algo-1}
For every matrix $C$, Algorithm \ref{alg:pivoting} returns a RAREF.
\end{proposition}

\begin{proof}
We check each of the RAREF properties:
\begin{itemize} \itemsep=0pt
\item[(a)] $L$ is non-singular as a product of elementary row-operation matrices. Because we apply the same operations to $L$ and to $T$, we have $L \cdot C = T$.
\item[(b)] The columns of $T$ indexed by $\bm{p}$ belong to the canonical base of $\FF_b^q$ thanks to step 3. They are distinct because at most one pivot is selected on each row.
\item[(c)] At the end of the algorithm, each row of $T$ contains exactly one pivot or is all-zero, hence the rank of $T$ equals the number of pivots.
\end{itemize}
\vskip-24pt
\end{proof}
}

\begin{algorithm}
\vskip 5pt
\modified{Let $L'$ = $L$, $T'$ = $T$  and $\bm{p}' = \bm{p}$.}
\begin{enumerate}
\item Locate a nonzero coefficient in the $i$th column of $L'$. Such a coefficient exists as $L'$ is non-singular. Say it is located in the $j$th row of $L'$ and it equals $a \in \FF_b$. Using a row permutation, exchange the $i$th and $j$th rows, both in $L'$ and $T'$. 
\item Using the nonzero coefficient $a$ at position $(i, i)$ of $L'$ and row transvections, zero all the other coefficients in the $i$th column of $L'$. Apply the same transvections to $T'$.  
\item \modified{Remove from $\bm{p}'$ the pivot in the $i$th row of $T'$.} Set to zero all the coefficients of the $i$th row of $L'$, except for $(i, i)$ which remains equal to $a$. The element $(i,i)$ of $L'$ is now completely isolated. Thus, it is now straightforward to update $T' = L' \cdot C$ into $T' = L' \cdot C'$: one has to replace the $i$th row of $T'$ by $a$ times the $i$th row of $C'$.
\item Pivot on the $i$th row by applying Algorithm \ref{alg:pivoting}.
\end{enumerate}
 Return \modified{$(L', T', \bm{p}')$}.
\caption{Updating the RAREF}
\label{alg:update}
\end{algorithm}

Given two composition matrices $C$ and $C'$ of size $q\times k$ which differ by exactly one row, we explain how to derive a RAREF \modified{$(L', T', \bm{p}')$ of $C'$} by reusing a known RAREF \modified{$(L, T, \bm{p})$ of $C$}. 
We can assume that $C$ has full rank, because otherwise $\cP$ would not be $(q_1, \dots q_s)$-equidistributed for the tuple $(q_1, \dots q_s)$ that corresponds to the composition matrix $C$.

\modified{As stated by Proposition \ref{prop:algo-2},} Algorithm \ref{alg:update}, whose operations are illustrated in 
Figure \ref{fig:exchange-gauss}, \modified{updates the RAREF so that all the properties listed in Definition \ref{def:almost-reduced-row-echelon-form} hold.}
The key advantage of this algorithm is that it requires only one row pivoting (step 4), 
whereas a full recomputation would require $q$ pivotings.

\modified{
\begin{proposition}
\label{prop:algo-2}
If $C$ and $C'$ are two matrices of the same shape that differ only by one row, $C$ has full rank, and $(L, T, \bm{p})$ is a RAREF of $C$, then the triplet $(L', T', \bm{p}')$ returned by Algorithm \ref{alg:update} is a RAREF of $C'$.
\end{proposition}

\begin{proof}
Let $i$ be the index of the row that differs between $C$ and $C'$. We check each of the RAREF properties:
\begin{itemize}\itemsep=0pt
	\item[(a)] Steps 1, 2 and 4 perform only row operations, so they preserve property (a). At step 3, the $i$th column of $L'$ is all-zero except for the $(i, i)$ coefficient.
	This ensures that replacing the $i$th row of $L'$ by $(0, \cdots, 0, a, 0, \cdots, 0)$ preserves the non-singularity of $L'$. It is then clear that replacing the $i$th row of $T'$ by $a c_i'$ implies $L' \cdot C' = T'$.
	\item[(b)] Row permutations leave the canonical base invariant so step 1 preserves property (b). 
	After step 2, the only column indexed by $\bm{p}'$ which may not belong to the canonical base must have its pivot on the $i$th row.
	As we remove this column from $\bm{p}'$ at step 3, property (b) is verified after step 3. Step 4 preserves property (b), due to step 3 of Algorithm \ref{alg:pivoting}. The distinctness stems from the fact that we may only add a single pivot on the $i$th row, which had no pivot before step 4.
	\item[(c)] Steps 1 and 2 change neither the rank of $T'$ nor the cardinality of $\bm{p}'$. 
	Because $C$ has full rank, the removal of the $i$th row in step 3 amounts to decreasing by 1 the number of pivots and the rank. 
	After adding back the new row, a new pivot is added during step 4 iff the row is linearly independent from the others (i.e., $C'$ has full rank). 
\end{itemize}
\vskip-24pt
\end{proof}
}

\begin{figure}
\footnotesize
\begin{footnotesize}
Step 1: 
Find a row with a nonzero coefficient $a$ in column $i$ of $L'$, and exchange this row 
 (blue row $j$) with row $i$, in $L'$ and in $T'$. 
\end{footnotesize}

\newcommand{\myunit}{0.89 cm}
\tikzset{
    node style sp/.style={draw,circle,minimum size=\myunit},
    node style ge/.style={circle,minimum size=\myunit}
}
\tikzset{BarreStyle/.style =   {opacity=.3,line width=5 mm,line cap=round,color=#1}}

\begin{changemargin}{-1.3cm}{-0.7cm} 

{\footnotesize
\begin{tikzpicture}[>=latex]
\tikzstyle{every node}=[font=\footnotesize]
\footnotesize
\matrix (L) [matrix of math nodes,%
             nodes = {node style ge},%
             left delimiter  = (,%
             right delimiter = )] at (-6.1*\myunit,0)
{%
  l_{11} & \cdots & l_{1i} & \cdots & l_{1q}  \\[-1em]
  \vdots & & \vdots & & \vdots  \\[-1em]
  l_{j1} & \ldots & |[node style sp]| a \neq 0 & \ldots & l_{jq}  \\[-1em]
  \vdots &  & \vdots &  & \vdots  \\[-1em]
  l_{i1} & \ldots & l_{ii} & \ldots & l_{iq} \\[-1em]
  \vdots &  & \vdots & & \vdots  \\[-1em]
  l_{q1} & \ldots & l_{qi} & \ldots & l_{qq} \\[-1em]
};
\node [draw,above=10pt] at (L.north) 
    { $L$ : $q$ rows, $q$ columns};
\draw [BarreStyle=blue!40] (L-3-1.west) to (L-3-5.east) ;
\draw [BarreStyle=red!50] (L-1-3.north) to (L-7-3.south) ;
\draw[<->, black] (L-3-1.west) to[bend right] (L-5-1.west);

\matrix (C) [matrix of math nodes,%
             nodes = {node style ge},%
             left delimiter  = (,%
             right delimiter =)] at (0,0)
{%
  c_{1} \\[-1em]
  \vdots \\[-1em]
  c_{i}  \\[-1em]
  \vdots \\[-1em]
  c_{q}  \\[-1em]
};
\node [draw,above=10pt] at (C.north) 
    { $C$ : $q$ rows $\{c_1, \dots, c_q\}$ };

\node [left=25pt] at (C.west) {$\times$};    
\node [right=40pt] at (C.east) {=};
    
\matrix (T) [matrix of math nodes,%
             nodes = {node style ge},%
             left delimiter  = (,%
             right delimiter = )] at (5.5*\myunit,0)
{%
 \vdots   \\[-1em]
  t_{j}  \\[-1em]
  \vdots  \\[-1em]
  t_{i}  \\[-1em]
  \vdots  \\[-1em]
};
\draw [BarreStyle=blue!40] (T-2-1.west) to (T-2-1.east) ;
\draw[<->, black] (T-2-1.east) to[bend left] (T-4-1.east);
\node [draw,above=10pt] at (T.north) 
    {$ T=L \modified{\cdot} C$ : $q$ rows $\{t_1, \dots, t_q\}$};
\end{tikzpicture}
}

\vspace{-0.2cm}
\begin{footnotesize}
Step 2: Use the nonzero element $a$ in $L'$ to zero all coefficients on column $i$ of $L'$, using transvections.
\end{footnotesize}

{\footnotesize
\begin{tikzpicture}[>=latex]
\tikzstyle{every node}=[font=\footnotesize]

\matrix (L) [matrix of math nodes,%
             nodes = {node style ge},%
             left delimiter  = (,%
             right delimiter = )] at (-6.1*\myunit,0)
{%
  l_{11} & \cdots & l_{1i} & \cdots & l_{1q}  \\[-1em]
  \vdots & & \vdots & & \vdots  \\[-1em]
  l_{i1} & \ldots &l_{ii}& \ldots & l_{iq}  \\[-1em]
  \vdots &  & \vdots &  & \vdots  \\[-1em]
  l_{j1} & \ldots & |[node style sp]| a & \ldots & l_{jq} \\[-1em]
  \vdots &  & \vdots & & \vdots  \\[-1em]
  l_{q1} & \ldots & l_{qi} & \ldots & l_{qq} \\[-1em]
};
\draw [BarreStyle=blue!40] (L-5-1.west) to (L-5-5.east) ;
\draw [BarreStyle=red!50] (L-1-3.north) to (L-7-3.south) ;
\draw[->, black] (L-5-3.west) to[bend left] (L-1-3.west);
\draw[->, black] (L-5-3.west) to[bend left] (L-3-3.west);
\draw[->, black] (L-5-3.west) to[bend right] (L-7-3.west);

\node [left=25pt] at (C.west) {$\times$};

\matrix (C) [matrix of math nodes,%
             nodes = {node style ge},%
             left delimiter  = (,%
             right delimiter =)] at (0,0)
{%
  c_{1} \\[-1em]
  \vdots \\[-1em]
  c_{i}  \\[-1em]
  \vdots \\[-1em]
  c_{q}  \\[-1em]
};
    
\node [right=40pt] at (C.east) {=};
    
\matrix (T) [matrix of math nodes,%
             nodes = {node style ge},%
             left delimiter  = (,%
             right delimiter = )] at (5.5*\myunit,0)
{%
  t_{1}  \\[-1em]
 \vdots   \\[-1em]
  t_{j}  \\[-1em]
  \vdots  \\[-1em]
  t_{i}  \\[-1em]
  \vdots  \\[-1em]
  t_{q}  \\[-1em]
};
\draw [BarreStyle=blue!40] (T-5-1.west) to (T-5-1.east) ;
\draw[->, black] (T-5-1.west) to[bend left] (T-1-1.west);
\draw[->, black] (T-5-1.west) to[bend left] (T-3-1.west);
\draw[->, black] (T-5-1.west) to[bend right] (T-7-1.west);
\end{tikzpicture}
}

\vspace{-0.2cm}
\begin{footnotesize}
Step 3: 
Set all coefficients of row $i$ of $L'$ to 0, except the diagonal coefficient.
\modified{Replace row $i$ of $T'$ with the new row $ac'_i$ (in green).}
Since the $i$th row of $C'$ (green row) interacts only with the green column of 
$L'$ in the matrix multiplication, \modified{we have $L' \cdot C' = T'$}. 

Step 4 (not shown): Pivot on row $i$ of $T'$ to obtain the RAREF $T'$.
\end{footnotesize}

\noindent
\begin{tikzpicture}[>=latex]
\tikzstyle{every node}=[font=\footnotesize]

\matrix (L) [matrix of math nodes,%
             nodes = {node style ge},%
             left delimiter  = (,%
             right delimiter = )] at (-6.2*\myunit,0)
{%
  l_{11} & \ldots & 0 & \ldots & l_{1q}  \\[-1em]
  \vdots &  & \vdots &  & \vdots  \\[-1em]
  0 &  0 & |[node style sp]| a & 0 & 0 \\[-1em]
  \vdots &  & \vdots &  & \vdots  \\[-1em]
  l_{q1} & \ldots & 0 & \ldots & l_{qq} \\[-1em]
};
\node [draw,below=10pt] at (L.south) 
    { $L'$ : $q$ rows, $q$ columns};
\draw [BarreStyle=blue!40] (L-3-1.west) to (L-3-5.east) ;
\draw [BarreStyle=green!50] (L-1-3.north) to (L-5-3.south) ;

\matrix (C) [matrix of math nodes,%
             nodes = {node style ge},%
             left delimiter  = (,%
             right delimiter =)] at (0,0)
{%
  c_{1} \\[-1em]
  \vdots \\[-1em]
  c'_{i}  \\[-1em]
  \vdots \\[-1em]
  c_{q}  \\[-1em]
};
\node [draw,below=10pt] at (C.south) 
    { $C'$ : $q$ rows, $k$ columns};
\draw [BarreStyle=green!50] (C-3-1.west) to (C-3-1.east) ;
    
\node [left=25pt] at (C.west) {$\times$};
\node [right=40pt] at (C.east) {=};
    
\matrix (T) [matrix of math nodes,%
             nodes = {node style ge},%
             left delimiter  = (,%
             right delimiter = )] at (5.5*\myunit,0)
{%
  t_{1} \\[-1em]
  \vdots \\[-1em]
  ac'_{i}  \\[-1em]
  \vdots \\[-1em]
  t_{q}  \\[-1em]
};
\draw [BarreStyle=green!50] (T-3-1.west) to (T-3-1.east) ;
\node [draw,below=10pt] at (T.south) 
    {$ T'=L'\modified{\cdot} C'$ : $q$ rows, $k$ columns};
\end{tikzpicture}

\end{changemargin}
\caption{Procedure to update a reduced almost row echelon form when row \modified{$i$} changes}
\label{fig:exchange-gauss}
\end{figure}

\subsection{Computing the t-values over many projections}
\label{subsec:many}

In this section, we extend our proposed methodology to the computation of the $t$-values of all or many projections. More precisely, given a vector $\cS$ of $s$ generator matrices of size $k \times k$, we aim at computing $t(\cS_{\frak u})$ 
(or equivalently $\rho(\cS_\mathfrak{u})$) for all projections $\frak u$ 
such that $|\frak u| \leq d_{\max}$. 
We propose a dynamic programming method that achieves this much more efficiently 
than computing the $t$-values independently across all the projections.
It exploits a recurrence that expresses $\rho(\cS_\mathfrak{u})$ in terms of 
the linear independence parameters of all strict subvectors $\cS_\mathfrak{v}$, 
with $\mathfrak{v}\subset \mathfrak{u}$.

We define the \emph{partial linear independence parameter} $\tilde{\rho}(\cS)$ of a vector of $d$ generator matrices $\cS$ 
as the maximum integer $q \geq d$ such that for every tuple of \emph{positive integers} $(q_1, ..., q_d)$ that sum to $q$, the composition matrix   
\[
 C =
\left[
\begin{array}{c}
  C_{1}^{(q_1)} \\
  \vdots \\
  C_{d}^{(q_d)}
\end{array}
\right]
\]
formed by 
the first $q_{1}$ rows of $C_{1}$, 
the first $q_{2}$ rows of $C_{2}$, \dots, 
the first $q_{d}$ rows of $C_{d}$, has full rank $q$. By convention, we set $\tilde{\rho}(\cS) = d - 1$ whenever there is no such $q$.
The difference with the linear independence parameter is that we consider here only tuples of strictly positive integers $(q_1, ..., q_d)$ and not non-negative integers as before. Note that
\[
  0 \leq {\rho}(\cS) \leq \tilde{\rho}(\cS) \leq k.
\]
With these ingredients, we can formulate the computation of $\rho(\cS)$ as a dynamic programming problem,
as follows.

\begin{proposition}
For each subset $\mathfrak{u} \subseteq \{1, \dots, s\}$  of cardinality greater than 2 
\begin{align} 
\rho(\cS_\mathfrak{u}) &= 
 \min \left(
   \tilde{\rho}(\cS_\mathfrak{u}),
   \min_{\emptyset \neq \mathfrak{v} \subset \mathfrak{u}} \rho(\cS_\mathfrak{v})
  \right) 
 = \min \left(
  \tilde{\rho}(\cS_\mathfrak{u}),\,
  \min_{j \in \mathfrak{u}} \rho(\cS_{\mathfrak{u}\setminus\{j\}})
\right).
\label{eq:dyn-prog}
\end{align}
\label{prop:rho-Su}
\end{proposition}

\begin{proof}
The first equality stems from the definition of $\rho$ and $\tilde{\rho}$,
the second equality comes from the fact that $\rho(\cS_\mathfrak{v'}) \geq \rho(\cS_\mathfrak{v})$
when $\mathfrak{v'} \subseteq \mathfrak{v}$.
\end{proof}

To compute $\rho(\cS_\mathfrak{u})$, for every $\mathfrak{u}$ such that $|\mathfrak{u}| \leq d_{\max}$,
we first compute $\rho(\cS_{\{j\}})$ for each $j \in\{1,\dots,s\}$. 
Thanks to the assumption of full projection regularity,
we already know that $\rho(\cS_{\{j\}}) = k$.
Then, we use the recurrence (\ref{eq:dyn-prog}) in the following manner: 
enumerate the subsets of $\{1, \dots, s\}$ by increasing cardinality until reaching cardinality $d_{\max}$. 
For each subset $\mathfrak{u}$, 
we first compute 
$q_{\max} = \min_{j \in \mathfrak{u}} \rho(\cS_{\mathfrak{u}\setminus\{j\}})$. We know that $\rho(\cS_\mathfrak{u}) \leq  q_{\max}$. If $q_{\max} < |\frak u|$, we know that  $\rho(\cS_\mathfrak{u}) =  q_{\max}$ as $\tilde{\rho}(\cS_{\frak u}) \geq |\frak u| - 1$. Otherwise, we test by decreasing values of $q \in \{q_{\max}, \dots, |\frak u|\}$ whether $\tilde{\rho}(\cS_\mathfrak{u}) \geq q$, 
by adapting the procedure described in the two previous subsections. The main technicality is that we need to modify the enumeration of composition matrices to stay in the subset of composition matrices which contain at least one row from each generator matrix of $\cS_{\frak u}$. 
To update the RAREF when the composition matrix changes by a single row,
we use the same method as before.

We visit the values of $q$ in decreasing order, from $q_{\max}$ to $|\frak u|$, because this
turns out to be faster when computing figures of merit that involve a large number 
of low-dimensional projections: 
for low-dimensional projections, the chances are that $t$ is small, so $\rho(\cS_\mathfrak{u})$ is nearer to $q_{\max}$ than to $|\frak u|$.
Another reason is that we implement an early stopping criterion: for a given $q$, as soon as we encounter a composition matrix which does not have full-rank, we stop and move on to $q-1$. This does not work in the increasing order, as we have to check that all matrices are full-rank for a given $q$ before moving on to $q+1$.

\subsection{Computation of the t-values for embedded nets}
\label{subsection:computation-embedded}

\hpl{Maybe this section should be expanded a bit, because it is an important 
 part of the contribution.}%
We now extend our proposed methodology to the case of embedded nets. 
More precisely, for given integers $1 \leq m_0 \leq k$ and a vector $\cS$ of $s$ generator matrices of size $k \times k$, 
we aim at computing $t(\cS_{\frak u, m})$ for all projections $\frak u$ such that 
$|\frak u| \leq d_{\max}$ and all $m_0 \leq m \leq k$.
\hpl{where $\cS_{\frak u, m}$ is the vector of generator matrices for the projection $\frak u$,
restricted to their first $m$ rows and $m$ columns.}

The procedure  
described in the \modified{paragraph that follows Proposition~\ref{prop:rho-Su}
can be adapted as follows.
For a given $q \in \{q_{\max}, \dots, |\frak u|\}$, let $l_q$ be the minimal number of columns  such that all the composition matrices of shape $q \times m$ for $l_q \leq m \leq k$ have full rank. 
By convention, if such a number does not exist, we set $l_q = k+1$.
The composition matrices of interest to compute $t(\cS_{\frak u, m})$ are the restrictions to their first $m$ columns of the composition matrices used to compute $t(\cS_{\frak u, k})$. 
Thus, for any $m \in \{m_0, \dots, k\}$, $\tilde{\rho}(\cS_{\frak u, m}) \geq q$ iff $m \geq l_q$. 
Thus computing $l_q$ for decreasing values of $q \in \{q_{\max}, \dots, |\frak u|\}$ yields the linear independence parameter for all $m$, hence solving the problem.

Algorithm \ref{alg:update} yields $l_q$ as the index of the rightmost column where a pivot is placed when cycling through all composition matrices, or $k+1$ if it is not possible to add a new pivot at some RAREF update. Let $l'_q$ be the latter quantity and let us prove that $l_q = l'_q$. First, $l_q = k+1$ iff $l'_q = k+1$, because all composition matrices have full rank ($l_q \leq k$) iff we add one pivot at each RAREF update ($l'_q \leq k$). Next we focus on the case where we achieve to add one pivot at each update. If a submatrix contains all the pivots, it has full rank, hence $l_q \leq l'_q$. During the update where a pivot was placed for the first time on column $l'_q$, because we choose the pivot as the \textit{first} nonzero coefficient of the new row, the restriction of the matrix to its first $l'_q - 1$ columns contains an all-zero row, so it does not have full rank. Hence we cannot have $l_q < l'_q$.
}
Thus, computing the figure of merit (\ref{eq:fig-merit-embedded}) 
for embedded nets does not add significant overhead compared to \modified{computing} the simpler figures (\ref{eq:fig-merit}) or (\ref{eq:fig-merit-infinity})\modified{, as we only have to keep tab on the right-most column where a pivot has been placed}.

\section{Complexity analysis}
\label{sec:complexity}

In this section, we perform a time complexity analysis of our method (MGL)
and compare it with methods S and PS from \cite{rSCH99a} and \cite{rPIR01a}.

For each method, we obtain formulas that provide bounds on the total number 
of vector additions in $\FF_b^k$ required to perform any of the following three computations,
given a vector $\cS$ of $s$ generator matrices of size $k \times k$:
(a) computing the $t$-value of the net, as described in 
Subsections~\ref{subsec:t-value} and \ref{subsection:gauss};
(b) computing the $t$-value over many projections, as described in 
Subsection~\ref{subsec:many}; and 
(c) computing the $t$-value over several projections for embedded nets,
as described in Subsection~\ref{subsection:computation-embedded}. 
For $b>2$, scalar multiplications are also used in some algorithms (PS and MGL). We do not count them as it would complicate the formulas without adding further insight
and because the number of scalar multiplications is normally less than the number of 
vector additions.

We also compare our method with the one of Dick and Matsumoto (DM) in \cite{rDIC13a}.
The latter uses weight enumerator polynomials to compute the $t$-value. 
The authors also outline an extension to compute the $t$-values over all
projections by using multivariate polynomials, but we do not consider 
this extension in this paper because we do not have an implementation and
it appears non-trivial to implement in practice.
Due to the different nature of their method, they express its complexity in terms of integer operations.

For $b=2$, if we assume that $k$ does not exceed the computer word length,
which is usually the case in the context of QMC, then vector additions in $\FF_b^k$ can be implemented as XOR operations between two integers. In this case, it makes sense to compare the complexity of DM to the others, because the elementary operation is the same.
However, for $b>2$, comparing vector additions in $\FF_b^k$ and integer operations does not really make sense. Moreover, vector additions must be computed as $k$ independent additions in $\FF_b$, so they are essentially $k$ times slower than for $b=2$. 

The memory space required by our method is not at all an issue,
as the most significant overhead is to store at each step two matrices of size $k \times k$.
Therefore we only study time complexity and not space complexity.

\subsection{Computing the $t$-value of the net}

We first look at the number of additions needed to compute or to update the RAREF:

\begin{proposition}
\label{prop:raref}
For general $b$ and a matrix of size $q \times k$, with $q \leq k$, computing the RAREF
by Algorithm \ref{alg:pivoting} requires at most $4 q^2$ vector additions,
whereas updating the RAREF by Algorithm \ref{alg:update} 
requires at most $6 q$ vector additions.
\end{proposition}

\begin{proof}
For Algorithm \ref{alg:pivoting}, at most $4q$ vector additions are done for each of the $q$ pivotings ($2q$ at step 1 and $2q$ at step 3), for a total of $4 q^2 $ operations.
For Algorithm \ref{alg:update}, $6q$ vector additions are needed  
($2q$ at step 2 and $4q$ at step 4), for a total of $6q$ operations.
\end{proof}

Next we bound the total number of vector additions required to compute a single $t$-value, 
for each method. For MGL, we give two bounds, depending on the enumerating order ($q$ increasing or $q$ decreasing), as explained in Subsection \ref{subsec:t-value}. We also provide simplified asymptotic expressions for the bounds
when $k$ tends to infinity with $s$ fixed.

\begin{proposition}
\label{prop:comp-t-value}
The number of operations to compute a $t$-value with the four methods 
considered is bounded as follows:

{\renewcommand{\arraystretch}{1.4}\rm
\begin{center}
\begin{tabular}{|c|c|}
  \hline
  S & ${\small\displaystyle \sum_{d=1}^{s} \sum_{q=d}^{k-t}} \binom{s}{d} \binom{q-1}{d-1} (b-1)^{d} b^{q-d}
			    =  \cO( (b+1)^{k-t+s-1}))$ \\
  PS  & ${\small\displaystyle \sum_{q=1}^{k-t}} \binom{q+s-1}{s-1} \binom{q}{2}  
		      =  \cO\left(\frac{(k-t)^{s+2}}{s!}\right)$ \\
  DM   & $\cO(k s b^k)$ \\
  MGL ($q$ decreasing) & ${\small\displaystyle \sum_{q=k-t}^k} \left[ 4q^2 + \left[ {{q+s-1}\choose{s-1}} -1 \right] 6q \right]   = \cO\left(\frac{k^{s+1}}{s!}\right)$ \\
  MGL ($q$ increasing) & ${\small\displaystyle \sum_{q=1}^{k-t}} \left[ 4q^2 + \left[ {{q+s-1}\choose{s-1}} -1 \right] 6q \right]   = \cO\left(\frac{(k-t)^{s+1}}{s!}\right)$ \\
  \hline
\end{tabular}
\end{center}
}
\end{proposition}

\begin{proof}
The upper bounds for S and PS are derived in \cite{rPIR01a}, in terms of the number
of vector additions.
The bound for DM can be found in \cite{rDIC13a}. 

To prove the bounds for MGL, we first note from Subsection \ref{subsec:t-value} that the number of composition matrices 
whose ranks are to be computed is at most $\sum_{q = k-t}^k \binom{q+s-1}{s-1}$ if we choose to visit the values of $q$ in decreasing order. 
For each $q \in \{k, \dots, k-t\}$, the first RAREF is computed entirely, 
followed by $\binom{q+s-1}{s-1} - 1$ updates, which takes 
$\sum_{q=k-t}^k \left[ 4q^2 + \left[ {{q+s-1}\choose{s-1}} -1 \right] 6q \right]$
vector additions in total, according to Proposition~\ref{prop:raref}.
By using the fact that $q \leq  {{q+s-1}\choose{s-1}}$ in the first inequality,
this expression simplifies as:
\begin{align*}
 \sum_{q=k-t}^k &\left[ 4q^2 + \left[ {{q+s-1}\choose{s-1}} -1 \right] 6q \right] \\
  &\leq 10  \sum_{q=k-t}^k {{q+s-1}\choose{s-1}} q 
	  ~=~ 10  \sum_{q=k-t}^k \left[ {{q+s}\choose{s}} - {{q+s-1}\choose{s}} \right] q \\
  &\leq 10  \sum_{q=k-t}^k \left[ {{q+s}\choose{s}}q - {{q+s-1}\choose{s}}(q-1) \right]  \\
   &= 10 k \binom{k+s}{s} - 10 (k-t-1) \binom{k-t-1+s}{s}.
 \end{align*}

If the values of $q$ are visited in increasing order, the computation above can be updated as follows:
$$
\sum_{q = 1}^{k-t} \left[ 4q^2 + \left[ {{q+s-1}\choose{s-1}} -1 \right] 6q \right]
\leq
10 (k-t) \binom{k-t+s}{s}
$$

The asymptotic expressions come from
\begin{equation*}
\label{eq:binom-inf}
\binom{k+s}{s} = \frac{(k+s)(k+s-1)...(k+1)}{s!} \sim \frac{k^s}{s!} 
  \text{ when } k \rightarrow \infty.
\end{equation*}

The above calculation can be adapted to prove the asymptotic expression for PS, which is tighter than the one given in \cite{rPIR01a}:
$$
\sum_{q=1}^{k-t} \binom{q+s-1}{s-1} \binom{q}{2}  
\leq (k-t)^2 \binom{k-t+s}{s} = \cO \left(\frac{(k-t)^{s+2}}{s!}\right).
$$
\vskip-24pt
\end{proof}

There is no uniform ranking between those methods, except that S is exponential in
both $s$ and $k$, DM is exponential only in $k$, and PS and MGL are exponential only in $s$. 
For MGL, the best choice in the enumerating order depends on the value of $t$: 
visiting the values of $q$ by decreasing order is the best choice when the $t$-value is small,
whereas increasing order wins when $t$ is far from 0.
For $b=2$ and typical values of $s$ and $k$, the threshold lies around $t=2$,
so the decreasing order wins only when $t$ is very close to 0.

\subsection{Computing the $t$-value over many projections}

We now compare the methods in terms of the computational effort to obtain the 
$t$-value over all projections over no more than $d_{\max}$ coordinates. 
This corresponds to the complexity of computing the figures of merit 
(\ref{eq:fig-merit})--(\ref{eq:fig-merit-infinity}).
Here the values of $q$ are enumerated by decreasing order, which is more effective
than increasing order when many low-order projections have to be examined.

\begin{proposition}	
\label{prop:complexity}
The number of vector additions to compute the $t$-value of all projections 
$\frak u$ such that $|\frak u| \leq d_{\max}$ is bounded as follows:
{\renewcommand{\arraystretch}{1.4}\rm
\begin{center}
\begin{tabular}{|c|c|}
  \hline
  S & $\sum_{d=1}^{d_{\max}} \sum_{q=d}^{k} \binom{s}{d} \binom{q-1}{d-1} (b-1)^{d} b^{q-d}$  \\
  PS & $k^2 \sum_{d=1}^{d_{\max}} \binom{s}{d} \binom{k+d}{d} $ \\
  DM & $\cO \left( kb^k \sum_{d=1}^{d_{\max}} \binom{s}{d} d \right)$   \\
	MGL &	 $ k \sum_{d=1}^{d_{\max}} \binom{s}{d} 
      \left[4k(k-d+1) +  6 \binom{k}{d} \right] $    \\
  \hline
\end{tabular}
\end{center}}
\end{proposition}

\begin{proof}
For S, the complexity is very similar to that in Proposition \ref{prop:comp-t-value}, because the algorithm to compute the $t$-value actually yields the $t$-value over all projections. The outer sum is restricted to the projections of cardinality $d\leq d_{\max}$.

For PS and DM, the complexity is the sum of the complexities of computing the $t$-values of each projection, which is respectively $k^2 \binom{k+d}{d}$ and $\cO(kdb^k)$ for a projection of size $d$ (by Proposition \ref{prop:comp-t-value}).

For MGL, for general $b$, the algorithm described in Subsection~\ref{subsec:many} requires at most 
\begin{equation}
\label{eq:compl-MGL}
\sum_{d=1}^{d_{\max}} \binom{s}{d} \sum_{q=d}^{k} \left[ 4 q^2 + \left[ \binom{q-1}{d-1} - 1 \right] 6 q \right]
\end{equation}
vector additions, because for each projection $\fraku$ of cardinality $d$, we compute by decreasing values of $q \in \{k, ..., d \}$ the ranks of all composition matrices.
As in Proposition \ref{prop:comp-t-value}, the cost of computing theses ranks is $4q^2 + \left[ \binom{q-1}{d-1} - 1 \right] 6q$, hence the formula (\ref{eq:compl-MGL}). 
It can be simplified as in the proof of Proposition \ref{prop:comp-t-value}:
\begin{eqnarray*}
  \sum_{q=d}^{k} \left[ 4q^2 + \left[ \binom{q-1}{d-1} - 1 \right] 6q \right] 
	&\leq&  4(k-d+1)k^2 + 6 \sum_{q=d}^{k} \binom{q-1}{d-1} q  \\
	&\leq&  4(k-d+1)k^2 + 6k \binom{k}{d}.
\end{eqnarray*}
\hmarion{
\begin{eqnarray*}
  &&  \sum_{q=d}^{k} \left[ 4q^2 + \left[ \binom{q-1}{d-1} - 1 \right] 6q \right] \\
  &\leq& 4(k-d+1)k^2 + 6 \sum_{q=d}^{k} \binom{q-1}{d-1} q \\
  & \leq & 4(k-d+1)k^2 + 6 \sum_{q=d}^{k} \left[ \binom{q}{d} q - \binom{q-1}{d} (q-1) \right] \\
  & = & 4(k-d+1)k^2 + 6k \binom{k}{d}.
\end{eqnarray*}
}
\vskip-24pt
\end{proof}

These expressions are difficult to compare in full generality for the parameters 
$d_{\max}$, $k$ and $s$. 
We can observe two benefits of MGL compared to PS: 
the algorithm described in Subsection~\ref{subsection:gauss} 
replaces the $k^2$ factor by $k$, and the algorithm described in Subsection~\ref{subsec:many} 
replaces the $\binom{k+d}{d}$ factor by approximately $k^2 +  \binom{k}{d}$.

We give simplified expressions for three choices of $d_{\max}$, namely
$d_{\max} = 2$, 3, and $s$.  Small values of $d_{\max}$ can be used when the computational
budget is limited and $s$ can be large, as in \cite{rJOE08a} for instance, where $d_{\max} = 2$ was used.
Taking $d_{\max} = s$ is the other extreme, and would be doable only if $s$ is small.

\begin{corollary} 
For the special cases of $d_{\max} = 2$, $d_{\max} = 3$, and $d_{\max} = s$,
the number of vector additions to compute the $t$-value of all projections 
$\frak u$ such that $|\frak u| \leq d_{\max}$ is bounded as follows:
{\renewcommand{\arraystretch}{1.4}\rm 
\begin{center}
\begin{tabular}{|c|c|c|c|c|}
  \hline
  & $d_{\max}=2$ & $d_{\max}=3$ & $d_{\max} = s$  \\
  \hline
  S & $\cO (s^2 b^k)$ & $\cO (s^3 k b^k)$ & $\sum_{d=1}^{s} \sum_{q=d}^{k} \binom{s}{d} \binom{q-1}{d-1} (b-1)^{d} b^{q-d}$ \\
  PS & $\cO (s^2 k^4)$ & $\cO (s^3 k^5)$ & $k^2 \sum_{d=1}^{s} \binom{s}{d}  {{k+d}\choose{d}}$ \\
  DM & $\cO (s^2 k b^k)$ & $\cO (s^3 k b^k)$ & $\cO (s 2^{s-1} kb^k)$ \\
  MGL & $\bm{\cO (s^2 k^3)}$ & $\bm{\cO (s^3 k^4)}$ & $ k \left[ 4 k^2 2^s + 6 \binom{s+k}{s} \right] $ \\
  \hline
\end{tabular}
\end{center}}
\end{corollary}

\begin{proof}
The expressions for $d_{\max}=2$ and $3$ follow from Theorem \ref{prop:complexity}.
For the case $d_{\max} = s$, for DM, we use the identity $\sum_{d=1}^{s} \binom{s}{d} d = s2^{s-1}$. For MGL, we use Vandermonde's convolution $\sum_{d=1}^{s} \binom{s}{d} \binom{k}{d} = \binom{s+k}{s}$. 
\end{proof}

We see that for $d_{\max} =2$ or 3, MGL has a better complexity \modified{than} all the other methods. 
For $d_{\max}=s$, the formulas are not easy to compare. 
If we take the limit as $k \rightarrow \infty$, we obtain similar asymptotic 
expressions as in Proposition \ref{prop:comp-t-value}.

\subsection{Computing the $t$-values for embedded nets}

Given two integers $m_0 \leq k$, we now consider the number of elementary operations to compute 
$t(\cS_{{\frak u},m})$ for all projections $\frak u$ such that $|\frak u| \leq d_{\max}$ 
and all $m_0 \leq m \leq k$. 
This represents the complexity of computing the figures of merit (\ref{eq:fig-merit-embedded}).
This may seem counter-intuitive, but for all four methods, 
considering embedded nets ($m_0 < k$) does not increase the workload compared to the 
non-embedded case ($m_0=k$). The reason is that the computations for $m = k$ include the 
computations for smaller values of $m$, by reordering some computations or 
saving intermediate results if needed. 
The details for MGL are given in Subsection~\ref{subsection:computation-embedded}.

\section{Numerical implementation and experiments}
\label{sec:numerical}
\label{sec:bench}

Here we compare the average CPU times required to compute the $t$-value of a random net,
and to compute the $t$-values of all projections of order up to $d_{\max}$ that are 
needed to obtain the figure of merit (\ref{eq:fig-merit}) or (\ref{eq:fig-merit-infinity}),
using our implementations of the four methods S, PS, DM, and MGL, 
discussed in the previous sections, for digital nets in base $b=2$.
All these methods are implemented efficiently in the \emph{LatNet Builder} 
software \cite{iLEC19l}.
All experiments were run on an Intel(R) Core(TM) i7-6500U CPU @ 2.50GHz processor.

\begin{figure}[h]
\begin{minipage}{.4\textwidth}
\begin{center}
\includegraphics[scale=0.6]{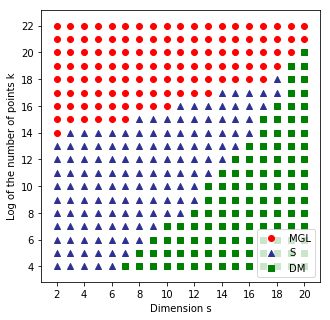}
\label{fig:dots}
\end{center}
\end{minipage}
\hspace{1cm}
\begin{minipage}{.5\textwidth}
\begin{center}
\includegraphics[height=0.85\textwidth,width=0.95\textwidth]{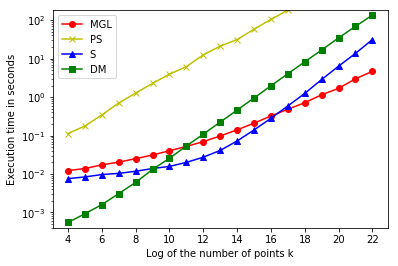}
\end{center}
\end{minipage}
\caption{Speed comparison for four methods (S, PS, DM, and MGL) to compute the $t$-value
   of a digital net in base 2 with $n = 2^k$ points, in $s$ dimensions.
  In the left panel, the symbol indicates which method is fastest as a function of $(s,k)$. 
	In the right panel, we fix $s=12$ and plot the execution time as a function of $k$, in log-log scale.}
\label{fig:benchmarking}
\end{figure}

\subsection{Computing the $t$-value of a random net}

Figure \ref{fig:benchmarking} compares the four methods 
in terms of the CPU time required to compute the 
$t$-value of a random fully projection-regular net.
For each considered pair $(s,k)$, between 20 and $10^5$ fully projection-regular nets 
were sampled uniformly, and the average CPU time to calculate the $t$-value was obtained. 
The number of samples was smaller for longer execution times.
In the left panel of Figure \ref{fig:benchmarking}, for each pair $(s,k)$, 
the symbol indicates which of the four methods is fastest on average, with our implementations.
As expected from Proposition \ref{prop:comp-t-value}, 
our strategy is the fastest for a large number of points in low dimension, 
while DM performs better in large dimension with a small number of points. 
Method S wins for easy cases (a small number of points in small dimension). 
In 19 or more dimensions, S is never the fastest method.
PS never wins, as it is always slower than MGL.

In the right panel, we fix $s=12$ and plot the execution time (in log scale) 
as a function of $k$.
We see that DM performs best for small point sets (when the required CPU time 
is under about 0.01 seconds),
S performs better for point sets of moderate size (CPU time between 0.01 and 0.5 seconds), 
and MGL performs better for large point sets (when it takes over 0.5 seconds).

\subsection{Computing the $t$-values of low-dimensional projections}

\begin{table}
\centering
\begin{tabular}{|ll|ccc|| ccc|}
\hline
         && \multicolumn{3}{c||}{S method} &  \multicolumn{3}{c|}{MGL method}  \\
\hline				
 $d_{\max}$ & $k$   & $s=5$  &  $s=20$ &  $s=100$  &  $s=5$  &  $s=20$ &  $s=100$  \\
\hline 
      & 10 & \textbf{0.4} & \textbf{4} & \textbf{76}
			     & 1  & 13 & 305\\
      & 15 & 34  & 58 & 1 522
			     & \textbf{2}  & \textbf{31} & \textbf{790}\\
  2   & 20 & 95  & 1 766 & 47 159
	         & \textbf{5} & \textbf{60} & \textbf{1 585}\\
      & 25 & 3 538 & 71 554 & > $10^6$
			     & \textbf{7} & \textbf{101} & \textbf{2 552}\\
      & 30 & 106 048 & > $10^6$ & > $10^6$
			     & \textbf{10} & \textbf{153} & \textbf{4 002}\\
\hline 
      & 10 & \textbf{0.6}  & \textbf{12} & \textbf{1 334}
			     & 1  & 45 & 5 023 \\
      & 15 & 4  & 221 & 25 832
			     & \textbf{3}  & \textbf{186} & \textbf{23 188} \\      
  3   & 20 & 147 & 7 869 & > $10^6$
	         & \textbf{8} & \textbf{556} & \textbf{70 215} \\
      & 25 & 5 495 & 306 162 & > $10^6$
			     & \textbf{17} & \textbf{1 286} & \textbf{168 941} \\
      & 30 & 182 193 & > $10^6$ & > $10^6$
			     & \textbf{30} & \textbf{2 478} &  \textbf{336 843} \\
\hline 
      & 10 & \textbf{0.5} & \textbf{81} & > $10^6$ & 1 & 237 & > $10^6$ \\
 5    & 15 & 5  & \textbf{1 110} & > $10^6$ & \textbf{4} & 2 013 & > $10^6$ \\
      & 20 & 168 & 59 586 & > $10^6$ & \textbf{12} & \textbf{16 936}  & > $10^6$ \\
\hline
\multicolumn{8}{c}{} \\
\hline
         && \multicolumn{3}{c||}{PS method} &  \multicolumn{3}{c|}{DM method}  \\
\hline				
 $d_{\max}$ & $k$   & $s=5$  &  $s=20$ &  $s=100$  &  $s=5$  &  $s=20$ &  $s=100$  \\
\hline 
      & 10 & 4 & 52 & 1 052
			     & 24  & 417 & 9 858 \\
      & 15 & 9  & 166 & 4 264
			     & 741  & 12 155 & 306 093\\
  2   & 20 & 26 & 492 & 12 607
	         & 25 440 & 413 915 & > $10^6$\\
      & 25 & 61 & 1 174 & 30 747
			     & 851 757 & > $10^6$ & > $10^6$\\
      & 30 & 139 & 2 432 & 63 433
			     & > $10^6$ & > $10^6$ & > $10^6$\\
\hline 
      & 10 & 7  & 509 & 66 444
			     & 62  & 4 647 & 589 036 \\
      & 15 & 30  & 3 127 & 406 101
			     & 1 865  & 140 408 & > $10^6$\\      
  3   & 20 & 130 & 12 280 & > $10^6$
	         & 63 412 & > $10^6$ & > $10^6$ \\
      & 25 & 390 & 38 414 & > $10^6$
			     & > $10^6$ & > $10^6$ & > $10^6$ \\
      & 30 & 1 011 & 99 078 & > $10^6$
			     & > $10^6$ & > $10^6$ & > $10^6$ \\
\hline 
      & 10 & 10 & 16 626 & > $10^6$ & 99 & 154 950 & > $10^6$ \\
5     & 15 & 64 & 175 731 & > $10^6$ & 3 180 & > $10^6$ & > $10^6$ \\
      & 20 & 335 & > $10^6$ & > $10^6$ & 106 845 & > $10^6$ & > $10^6$ \\
\hline
\end{tabular}
\caption{Average CPU times (in milliseconds) for computing the figure of merit (\ref{eq:fig-merit-infinity}) 
  for $d_{\max} = 2$, 3 and 5, various $s$ and $k$, with methods S, PS, DM, and MGL.}
\label{tab:finite-order-benchmarking}
\end{table}

For large high-dimensional nets, e.g., if $s \geq 20$ and $k \geq 20$, 
looking at the $t$-values of low-dimensional projections is more reasonable 
and also more relevant than computing the $t$-value of the full net \cite{rJOE08a,vLEC09f}.
For $s=k=20$, for example, it takes 100 seconds to compute the $t$-value 
of the full net and only \modified{0.5} second to compute the figure of merit 
(\ref{eq:fig-merit-infinity}) for $d_{\max} = 3$.
Table~\ref{tab:finite-order-benchmarking} compares all four methods
in terms of the CPU time required to compute the figure of merit (\ref{eq:fig-merit-infinity})
with $m=k$, $\tilde{t}(|\mathfrak{u}|, m, t) = t$ (meaning we take the maximum of the $t$-values of each projections), 
and weights $\gamma_{\mathfrak{u}} = 1$ for $|\mathfrak{u}| \leq d_{\max}$ and 0 otherwise.
We look at projections up to order $d_{\max} = 2$, 3 and 5, number of dimensions $s = 5$, 20, 100,
and number of points $n$ from $2^{10}$ to $2^{30}$.
Each experiment was repeated for a sample of 20 random independent uniformly-distributed 
fully projection-regular nets and we report the average CPU times.
For each combination of $d_{\max}$, $s$, and $k$, the smallest value is in bold.
For the entries marked ``$> 10^6$'', 1000 seconds were not sufficient on average.
We see from the table that MGL performs best in all cases, 
except when the number of points is small ($n=2^{10}$), in which case S performs better. 
For such a small number, the execution time is very small anyway.
For large $s$ and $n$ ($n\geq 2^{25}$), MGL is more than ten times quicker than the second best algorithm (PS).

\section{Conclusion}

Being able to compute $t$-values efficiently is critical for the practitioner who wants to 
find quickly (in real time) good digital nets with arbitrary values of $s$ and $k$, arbitrary weights, etc.,
and for the researcher who wants to study numerically the properties of figures of merit based on $t$-values.
We offer in this paper a new efficient algorithm to compute the $t$-value of a net and of its
projections, and to do so for a series of embedded nets as well.
We compare the performance of our algorithm with three other ones, 
proposed in \cite{rSCH99a}, \cite{rDIC13a} and \cite{rPIR01a}. 
We also \modified{implement} these four algorithms in the open-source software 
\emph{LatNet Builder} \cite{iLEC19l} for the most commonly used base, $b=2$.

According to our numerical comparisons, the method of \cite{rSCH99a} is the best performer
when $s$ and $k$ are small, the method of \cite{rDIC13a} performs better when $s$ is large and $k$ is small,
and our new method is the best performer when $k$ is large and either $s$ is small or we consider 
only the projections of relatively low order.  
The latter case is very relevant for practical applications.
For example, we can easily compute a figure of merit that accounts for the $t$-values of all 
projections of order 3 or less for a net with $2^{25}$ points in $s=100$ dimensions, 
whereas with all others methods, the cost of doing  this is prohibitive.

\section{References}

\bibliographystyle{model1b-num-names}
\begingroup
\renewcommand{\section}[2]{}%
\bibliography{vrt,random,math,stat,ift,optim,simul}
\endgroup
\end{document}